\documentstyle[12pt,amssym]{article}
\font\sixteenof=msbm10 at 16pt

\def\C{\mbox{$\Bbb C$}}
\def\R{\mbox{$\Bbb R$}}
\def\N{\mbox{$\Bbb N$}}

\def\case#1#2{{\textstyle{#1\over #2}}}
\def\sech{\mathop{\rm sech}\nolimits}
\def\cosech{\mathop{\rm cosech}\nolimits}

\title{
Non-Hermitian Hamiltonians with real and complex eigenvalues in a Lie-algebraic
framework}

\author{B.\ Bagchi $^{a,}$\thanks{E-mail: bbagchi@cucc.ernet.in}\ , C.\ Quesne
$^{b,}$\thanks{Directeur de recherches FNRS; E-mail: cquesne@ulb.ac.be} \\
{\small \sl $^a$ Department of Applied Mathematics, University of Calcutta,} \\
{\small \sl 92 Acharya Prafulla Chandra Road, Calcutta 700 009, India}\\
{\small \sl $^b$ Physique Nucl\'eaire Th\'eorique et Physique
Math\'ematique,  Universit\'e Libre de Bruxelles,} \\ {\small \sl Campus de la
Plaine CP229, Boulevard~du Triomphe, B-1050 Brussels, Belgium}}
\date{ }
\begin{document}
\baselineskip=22pt plus 1pt minus 1pt
%%%%%%%%%%%%%%%%%%%%%%%%%%%%%%%%%%%%%%%%%%%%%%%%%%%%%%%%%%
\maketitle

\begin{abstract}
We show that complex Lie algebras (in particular sl(2,\C)) provide us with an elegant
method for studying the transition from real to complex eigenvalues of a class of
non-Hermitian Hamiltonians: complexified Scarf II, generalized P\"oschl-Teller, and
Morse. The characterizations of these Hamiltonians under the so-called
pseudo-Hermiticity are also discussed.
\end{abstract}

\vspace{0.5cm}



\bigskip\noindent
Corresponding author: C.\ Quesne, Physique Nucl\'eaire Th\'eorique et Physique
Math\'e\-ma\-ti\-que,  Universit\'e Libre de Bruxelles, Campus de la Plaine
CP229, Boulevard du Triomphe, B-1050 Brussels, Belgium

\noindent
Telephone: 32-2-6505559

\noindent
Fax: 32-2-6505045

\noindent
E-mail: cquesne@ulb.ac.be 
\newpage
%
%========================================================================
%
\section{Introduction}

Some years ago, it was suggested~\cite{bender98a} that PT symmetry might be
responsible for some non-Hermitian Hamiltonians to preserve the reality of their
bound-state eigenvalues provided it is not spontaneously broken, in which case their
complex eigenvalues should come in conjugate pairs. Following this, several non-Hermitian
Hamiltonians (including the non-PT-symmetric ones~\cite{cannata, khare, bagchi00a})
with real or complex spectra have been analyzed using a variety of techniques, such as
perturbation theory, semiclassical estimates, numerical experiments, analytical arguments,
and algebraic methods. Among the latter, one may quote those connected with
supersymmetrization~\cite{cannata, andrianov, bagchi00b, znojil00, levai00, bagchi01a,
mosta02a}, or some generalizations thereof~\cite{bagchi02},
quasi-solvability~\cite{khare, bender98b, znojil99, bagchi00c, bagchi01b, kaushal}, and
potential algebras~\cite{bagchi00a, levai01a}.\par
%
%------------------------------------------------------------------------------------------------------
%
Recently, it has been shown that under some rather mild assumptions, the existence of
real or complex-conjugate pairs of eigenvalues can be associated with a class of
non-Hermitian Hamiltonians distinguished by either their so-called (weak) {\em
pseudo-Hermiticity\/} [i.e., such that $\eta H \eta^{-1} = H^{\dagger}$, where $\eta$ is
some (Hermitian) linear automorphism] or their invariance under some antilinear
operator~\cite{mosta02b, solombrino}. In such a context, pseudo-Hermiticity under
imaginary shift of the coordinate has been identified as the explanation of the occurrence
of real or complex-conjugate eigenvalues for some non-PT-symmetric
Hamiltonians~\cite{ahmed01a}.\par
%
%------------------------------------------------------------------------------------------------------------
%
In the course of time, there has been a growing interest in determining the critical
strengths of the interaction at which PT symmetry (or some generalization) becomes
spontaneously broken, i.e., they appear {\em regular\/} complex-energy solutions, where
by regular we mean eigenfunctions satisfying the asymptotic boundary conditions
$\psi(\pm \infty) \to 0$, so that they are normalizable in a generalized
sense~\cite{mosta02b, ahmed01a, znojil01a, ahmed01b}. Some analytical results have
been obtained both for PT-symmetric potentials~\cite{ahmed01b, znojil01b, levai01b,
znojil01c} and for potentials that are pseudo-Hermitian under imaginary shift of the
coordinate~\cite{ahmed01a}.\par
%
%-----------------------------------------------------------------------------------------------------
% 
In the present Letter, we wish to show that complex Lie algebras provide us with an easy
and elegant method for studying the transition from real to complex eigenvalues,
corresponding to {\em regular\/} eigenfunctions, of (PT-symmetric or
non-PT-symmetric) pseudo-Hermitian and non-pseudo-Hermitian Hamiltonians.\par
%
%============================================================
%
\section{\boldmath Non-Hermitian Hamiltonians in an sl(2, {\sixteenof C}) framework}

The generators $J_0$, $J_+$, $J_-$ of the complex Lie algebra sl(2,\C), characterized
by the commutation relations
\begin{equation}
  [J_0, J_{\pm}] = \pm J_{\pm}, \qquad [J_+, J_-] = - 2J_0,  
\end{equation}
can be realized as differential operators~\cite{bagchi00a}
\begin{equation}
  J_0 = - {\rm i} \frac{\partial}{\partial\phi}, \qquad J_{\pm} = e^{\pm{\rm i}\phi}
  \left[\pm\frac{\partial}{\partial x} + \left({\rm i} \frac{\partial}{\partial\phi}
  \mp \frac{1}{2}\right) F(x) + G(x)\right],  \label{eq:J}  
\end{equation}
depending upon a real variable $x$ and an auxiliary variable $\phi \in [0, 2\pi)$,
provided the two complex-valued functions $F(x)$ and $G(x)$ in (\ref{eq:J}) satisfy
coupled differential equations
\begin{equation}
  F' = 1 - F^2, \qquad G' = - FG.  \label{eq:F-G-diff}  
\end{equation}
Here a prime denotes derivative with respect to spatial variable $x$.\par
%
%------------------------------------------------------------------------------------------------------
% 
The solutions of Eq.~(\ref{eq:F-G-diff}) fall into the following three classes:
\begin{equation}
\begin{array}{lll}
      {\rm I}: & F(x) = \tanh(x-c-{\rm i}\gamma), & G(x) = (b_R+{\rm i}b_I)
            \sech(x-c-{\rm i}\gamma), \\[0.2cm]
      {\rm II}: & F(x) = \coth(x-c-{\rm i}\gamma), & G(x) = (b_R+{\rm i}b_I)
            \cosech(x-c-{\rm i}\gamma), \\[0.2cm]
      {\rm III}: & F(x) = \pm 1, & G(x) = (b_R+{\rm i}b_I) e^{\mp x},  
\end{array} \label{eq:F-G}  
\end{equation}
where $c$, $b_R$, $b_I \in \R$ and $-\frac{\pi}{4} \le \gamma < \frac{\pi}{4}$, thus
providing us with three different realizations of sl(2,\C). For $b_I = \gamma = 0$, the
latter reduce to corresponding realizations of ${\rm sl(2,\R)} \simeq {\rm so(2,1)}$, for
which $J_0 = J_0^{\dagger}$ and $J_- = J_+^{\dagger}$~\cite{englefield}.\par
%
%-----------------------------------------------------------------------------------------------------
%
The sl(2, \C) Casimir operator corresponding to the differential realizations of
type~(\ref{eq:J}) can be written as
\begin{eqnarray}
  J^2 & \equiv & J_0^2 \mp J_0 - J_{\pm} J_{\mp} \nonumber \\
  & = & \frac{\partial^2}{\partial x^2} - \left(\frac{\partial^2}{\partial \phi^2} +
  \frac{1}{4}\right) F' + 2 {\rm i} \frac{\partial}{\partial \phi} G' - G^2 - \frac{1}{4}. 
\end{eqnarray}
\par
%
%---------------------------------------------------------------------
%
In this work, we are going to consider the sl(2, \C) irreducible representations spanned by
the states
\begin{equation}
  |km\rangle = \Psi_{km}(x, \phi) = \psi_{km}(x) \frac{e^{{\rm i} m\phi}}{\sqrt{2\pi}}
  \label{eq:Psi}
\end{equation}
with fixed $k$, for which
\begin{equation}
  J_0 |km\rangle = m |km\rangle, \qquad J^2 |km\rangle = k(k-1) |km\rangle, 
  \label{eq:irrep}
\end{equation}
and
\begin{equation}
  k = k_R + {\rm i}k_I, \qquad m = m_R + {\rm i}m_I, \qquad m_R = k_R + n, \qquad 
  m_I = k_I, 
\end{equation}
where $k_R$, $k_I$, $m_R$, $m_I \in \R$ and $n \in \N$. The states with $m=k$ or
$n=0$ satisfy the equation $J_- |k k\rangle = 0$, while those with higher values of $m$
(or $n$) can be obtained from them by repeated applications of $J_+$ and use of the
relation $J_+ |k m\rangle \propto |k\, m+1\rangle$.\par
%
%----------------------------------------------------------------------------------------------
%
When the parameter $m$ is real, i.e., $m_I = 0$, we can get rid of the auxiliary variable
$\phi$ by extending the definition of the pseudo-norm with a multiplicative integral over
$\phi$ from 0 to $2\pi$. In the case $m$ is complex, i.e., $m_I \ne 0$, a similar result can
be obtained through an appropriate change of the integral over $\phi$. In the former
(resp.\ latter) case, $J_0$ is a Hermitian (resp.\ non-Hermitian) operator.\par
%
%------------------------------------------------------------------------------------------------------
%
{}From the second relation in Eq.~(\ref{eq:irrep}), it follows that the functions
$\psi_{km}(x)$ of Eq.~(\ref{eq:Psi}) obey the Schr\"odinger equation
\begin{equation}
  - \psi''_{km} + V_m \psi_{km} = - \left(k - \case{1}{2}\right)^2 \psi_{km},
  \label{eq:SE}
\end{equation}
where the family of potentials $V_m$ is defined by
\begin{equation}
  V_m = \left(\case{1}{4} - m^2\right) F' + 2m G' + G^2.  
\end{equation}
Since the irreducible representations of sl(2, \C) correspond to a given eigenvalue in
Eq.~(\ref{eq:SE}) and the corresponding basis states to various potentials $V_m$, $m =
k$, $k+1$, $k+2$,~\ldots, it is clear that sl(2, \C) is a potential algebra for the family of
potentials $V_m$ (see~\cite{englefield} and references quoted therein).\par 
%
%---------------------------------------------------------------------------------------------------
% 
To the three classes of solutions of Eq.~(\ref{eq:F-G-diff}), given in Eq.~(\ref{eq:F-G}),
we can now associate three classes of potentials:
\begin{eqnarray}
  {\rm I:\quad} V_m & = & \left[(b_R + {\rm i}b_I)^2 - (m_R + {\rm i}m_I)^2 +
         \case{1}{4}\right] \sech^2\tau \nonumber \\ 
  && \mbox{} - 2 (m_R + {\rm i}m_I)(b_R + {\rm i}b_I) \sech\tau \tanh\tau, \qquad 
         \tau = x - c - {\rm i}\gamma,  \label{eq:VmI} \\
  {\rm II:\quad} V_m & = & \left[(b_R + {\rm i}b_I)^2 + (m_R + {\rm i}m_I)^2 -
         \case{1}{4}\right] \cosech^2\tau \nonumber \\
  && \mbox{} - 2(m_R + {\rm i}m_I) (b_R + {\rm i}b_I) \cosech\tau \coth\tau, \qquad 
         \tau = x - c - {\rm i}\gamma,  \label{eq:VmII} \\         
  {\rm III:\quad} V_m & = & (b_R + {\rm i}b_I)^2 e^{\mp 2x} \mp 2(m_R + {\rm i}m_I)
         (b_R + {\rm i}b_I) e^{\mp x}.  \label{eq:VmIII} 
\end{eqnarray}
It is worth stressing that in the generic case, such complex potentials are not invariant
under PT symmetry.\par
%
%-------------------------------------------------------------------------------------------------------
%
Equation~(\ref{eq:SE}) can also be rewritten as
\begin{equation}
  - \psi^{(m)\prime\prime}_n + V_m \psi^{(m)}_n = E^{(m)}_n \psi^{(m)}_n,
  \label{eq:SE-bis}
\end{equation}
with $\psi_{km}(x) = \psi^{(m)}_n(x)$ and
\begin{equation}
  E^{(m)}_n = - \left(m_R + {\rm i}m_I - n - \case{1}{2}\right)^2.  \label{eq:E}
\end{equation}
Real (resp.\ complex) eigenvalues therefore correspond to $m_I = 0$ (resp.\ $m_I \ne
0$).\par
%
%---------------------------------------------------------------------------------------------------------
%
To be acceptable solutions of Eq.~(\ref{eq:SE-bis}), the functions $\psi^{(m)}_n(x)$
have to be regular, i.e., such that $\psi^{(m)}_n(\pm\infty) \to 0$. It is straightforward
to determine under which conditions there exist acceptable solutions of
Eq.~(\ref{eq:SE-bis}) with $n=0$. The functions $\psi^{(m)}_0(x)$ are indeed easily
obtained by solving the first-order differential equation $J_- \Psi_{mm}(x, \phi) = 0$. For
the three classes of potentials (\ref{eq:VmI}) -- (\ref{eq:VmIII}), the results read
\begin{eqnarray}
  {\rm I:\quad} \psi^{(m)}_0(x) & \propto & (\sech\tau)^{m_R + {\rm i}m_I - 1/2}
         \exp[(b_R + {\rm i}b_I) \arctan(\sinh\tau)],  \label{eq:psiI} \\
  {\rm II:\quad} \psi^{(m)}_0(x) & \propto & (\sinh\case{\tau}{2})^{b_R + {\rm i}b_I -
         m_R - {\rm i}m_I + 1/2} (\cosh\case{\tau}{2})^{- b_R -
         {\rm i}b_I - m_R - {\rm i}m_I + 1/2}, \\      
  {\rm III:\quad} \psi^{(m)}_0(x) & \propto & \exp[-(m_R + {\rm i}m_I - \case{1}{2})
         x - (b_R + {\rm i}b_I) e^{-x}].  \label{eq:psiIII}
\end{eqnarray}
Such functions are regular provided $m_R  > \frac{1}{2}$ and $b_R  > 0$, where the
second condition applies only to class~III.\par
%
%-----------------------------------------------------------------------------------------------------------
%
In the remainder of this letter, we shall illustrate the general theory developed in the
present section with some selected examples.\par
%
%===========================================================
%
\section{Complexified Scarf II potential}

The potential
\begin{equation}
  V(x) = - V_1 \sech^2 x - {\rm i} V_2 \sech x \tanh x, \qquad V_1 > 0, \qquad
  V_2 \ne 0,  \label{eq:Scarf}
\end{equation}
which belongs to class~I defined in Eq.~(\ref{eq:VmI}), is a complexification of the real
Scarf II potential~\cite{cooper}. It is not only invariant under PT symmetry but also
P-pseudo-Hermitian. Comparison between Eqs.~(\ref{eq:VmI}) and (\ref{eq:Scarf}) shows
that it corresponds to $c = \gamma = 0$ and
\begin{eqnarray}
  b_R^2 - b_I^2 - m_R^2 + m_I^2 + \case{1}{4} & = & - V_1, \label{eq:Scarf-eq1} \\
  b_R b_I - m_R m_I & = & 0, \label{eq:Scarf-eq2} \\
  m_R b_R - m_I b_I & = & 0, \label{eq:Scarf-eq3} \\
  2(m_R b_I + m_I b_R)  & = & V_2, \label{eq:Scarf-eq4}  
\end{eqnarray}
where we may assume $b_I \ne 0$ since otherwise the sl(2, \C) generators~(\ref{eq:J})
would reduce to sl(2, \R) ones.\par
%
%-------------------------------------------------------------------------------------------------------
%
To be able to apply the results of the previous section, the only thing we have to do is to
solve Eqs.~(\ref{eq:Scarf-eq1}) -- (\ref{eq:Scarf-eq4}) in order to express the sl(2, \C)
parameters $b_R$, $b_I$, $m_R$, $m_I$ in terms of the potential parameters $V_1$,
$V_2$. Equations~(\ref{eq:Scarf-eq3}) and (\ref{eq:Scarf-eq4}) yield
\begin{equation}
  m_R = \frac{V_2 b_I}{2(b_R^2 + b_I^2)}, \qquad m_I = \frac{V_2 b_R}{2(b_R^2 +
  b_I^2)}.  \label{eq:Scarf-inter1} 
\end{equation}
On inserting these results into Eqs.~(\ref{eq:Scarf-eq1}) and (\ref{eq:Scarf-eq2}), we get
the relations
\begin{eqnarray}
  (b_R^2 - b_I^2) \left(1 + \frac{V_2^2}{4(b_R^2 + b_I^2)^2}\right) & = & - V_1 - 
        \frac{1}{4}, \label{eq:Scarf-eq1-bis} \\
  b_R b_I \left(1 - \frac{V_2^2}{4(b_R^2 + b_I^2)^2}\right) & = & 0.     
\end{eqnarray}
The latter is satisfied if either $b_R = 0$ or $b_R \ne 0$ and $b_R^2 + b_I^2 =
\frac{1}{2} |V_2|$. It now remains to solve Eq.~(\ref{eq:Scarf-eq1-bis}) in those two
possible cases.\par
%
%--------------------------------------------------------------------------------------------------------
%
If we choose $b_R = 0$, then Eq.~(\ref{eq:Scarf-eq1-bis}) reduces to a quadratic
equation for $b_I^2$, which has real positive solutions
\begin{equation}
  b_I^2 = \case{1}{4} \left(\sqrt{V_1 + \case{1}{4} + V_2} + \epsilon_I \sqrt{V_1 +
  \case{1}{4} - V_2}\right)^2, \qquad \epsilon_I = \pm 1, \label{eq:Scarf-inter2} 
\end{equation}
provided $|V_2| \le V_1 + \frac{1}{4}$. Equation (\ref{eq:Scarf-inter2}) then yields for
$b_I$ the possible solutions 
\begin{equation}
  b_I = \case{1}{2} \epsilon'_I \left(\sqrt{V_1 + \case{1}{4} + V_2} + \epsilon_I
  \sqrt{V_1 + \case{1}{4} - V_2}\right), \qquad \epsilon_I, \epsilon'_I = \pm 1,
\end{equation}
while Eq.~(\ref{eq:Scarf-inter1}) leads to $m_R = V_2/(2b_I)$ and $m_I = 0$.\par
%
%---------------------------------------------------------------------------------------------------------
%
{}From the regularity condition $m_R > \frac{1}{2}$ of $\psi^{(m)}_0(x)$, given in
Eq.~(\ref{eq:psiI}), it then follows that $b_I$ must have the same sign as $V_2$, which
we denote by $\nu$. Furthermore, we must choose $\epsilon'_I = +1$ or $\epsilon'_I = -
\epsilon_I$ according to whether $\nu = +1$ or $\nu = -1$.\par
%
%------------------------------------------------------------------------------------------------------
%
The first set of solutions of Eqs.~(\ref{eq:Scarf-eq1}) -- (\ref{eq:Scarf-eq4}), compatible
with the regularity condition of $\psi^{(m)}_0(x)$, is therefore given by
\begin{eqnarray}
  b_R & = & 0, \qquad b_I = \case{1}{2} \nu \left(\sqrt{V_1 + \case{1}{4} + |V_2|} -
          \epsilon \sqrt{V_1 + \case{1}{4} - |V_2|}\right), \nonumber \\
  m_R & = & \case{1}{2} \left(\sqrt{V_1 + \case{1}{4} + |V_2|} + \epsilon \sqrt{V_1 +
          \case{1}{4} - |V_2|}\right), \qquad m_I = 0, \qquad \epsilon = \pm 1,
\end{eqnarray}
where $\epsilon = - \epsilon_I$, provided $|V_2| \le V_1 + \frac{1}{4}$ and 
$\sqrt{V_1 + \frac{1}{4} + |V_2|} + \epsilon \sqrt{V_1 + \frac{1}{4} - |V_2|} > 1$.\par
%
%------------------------------------------------------------------------------------------------------
%
On inserting these results into Eq.~(\ref{eq:E}), we get two series of real eigenvalues
\begin{equation}
  E_{n,\epsilon} = - \left[\case{1}{2} \left(\sqrt{V_1 + \case{1}{4} + |V_2|} + \epsilon 
  \sqrt{V_1 + \case{1}{4} - |V_2|}\right) - n - \case{1}{2}\right]^2, \qquad \epsilon =
  \pm 1.  \label{eq:Scarf-E1}
\end{equation}
By studying the regularity condition of the associated eigenfunctions obtained by
successive applications of $J_+$ on $\psi^{(m)}_0(x)$, it can be shown that $n$ is
restricted to the range $n= 0$, 1, 2,~\ldots ${}< \frac{1}{2} \left(\sqrt{V_1 + 
\frac{1}{4} + |V_2|} + \epsilon \sqrt{V_1 + \frac{1}{4} - |V_2|} - 1\right)$.\par
%
%------------------------------------------------------------------------------------------------------
% 
If, on the contrary, we choose $b_R \ne 0$ and $b_R^2 + b_I^2 = \case{1}{2} |V_2|$,
then Eq.~(\ref{eq:Scarf-eq1-bis}) leads to $b_R^2 - b_I^2 = - \frac{1}{2}(V_1 +
\frac{1}{4})$, so that
\begin{equation}
  b_R = \case{1}{2} \epsilon_R \sqrt{|V_2| - V_1 - \case{1}{4}}, \qquad b_I =
  \case{1}{2} \epsilon_I \sqrt{|V_2| + V_1 + \case{1}{4}}, \qquad \epsilon_R, \epsilon_I
  = \pm 1, 
\end{equation}
provided $|V_2| > V_1 + \frac{1}{4}$.\par
%
%------------------------------------------------------------------------------------------------------
%
On inserting such results into Eq.~(\ref{eq:Scarf-inter1}) and imposing the regularity
condition $m_R > \frac{1}{2}$, we obtain $\epsilon = \nu$. The second set of solutions
of Eqs.~(\ref{eq:Scarf-eq1}) -- (\ref{eq:Scarf-eq4}), compatible with the regularity
condition of $\psi^{(m)}_0(x)$, is therefore given by
\begin{eqnarray}
  b_R  & = & \case{1}{2} \nu \epsilon \sqrt{|V_2| - V_1 - \case{1}{4}}, \qquad b_I =
         \case{1}{2} \nu \sqrt{|V_2| + V_1 + \case{1}{4}}, \nonumber \\
  m_R  & = & \case{1}{2} \sqrt{|V_2| + V_1 + \case{1}{4}}, \qquad m_I =
         \case{1}{2} \epsilon \sqrt{|V_2| - V_1 - \case{1}{4}},  \qquad \epsilon = \pm 1, 
\end{eqnarray} 
where we have set $\epsilon = \nu \epsilon_R$. Here we must assume $|V_2| > V_1 +
\frac{1}{4}$ and $|V_2| + V_1 + \frac{1}{4} > 1$.\par
%
%-------------------------------------------------------------------------------------------------------
%
This set of solutions is associated with a series of complex-conjugate pairs of eigenvalues
\begin{equation}
  E_{n,\epsilon} = - \left[\case{1}{2} \left(\sqrt{|V_2| + V_1 + \case{1}{4}} + {\rm i}
  \epsilon \sqrt{|V_2| - V_1 - \case{1}{4}}\right) - n - \case{1}{2}\right]^2, \qquad
  \epsilon = \pm 1,  \label{eq:Scarf-E2}
\end{equation}
where it can be shown that $n$ varies in the range $n = 0$, 1, 2,~\ldots ${}<
\case{1}{2} \left(\sqrt{|V_2| + V_1 + \case{1}{4}} - 1\right)$.\par
%
%-------------------------------------------------------------------------------------------------------
%
We conclude that for increasing values of $|V_2|$, the two series of real eigenvalues
(\ref{eq:Scarf-E1}) merge when $|V_2|$ reaches the value $V_1 + \frac{1}{4}$, then
disappear while complex-conjugate pairs of eigenvalues (\ref{eq:Scarf-E2}) make their
appearance, as already found elsewhere by another method~\cite{ahmed01b}. Had we
chosen the parametrization $V_1 = B^2 + A(A+1)$, $V_2 = - B(2A+1)$, with
$A$ and $B$ real, as we did in Ref.~\cite{bagchi00a}, we would obtain that the
condition $|V_2| \le V_1 + \frac{1}{4}$ is always satisfied, thus only getting the two
series of real eigenvalues (\ref{eq:Scarf-E1}).\par
%
%=============================================================
% 
\section{Complexified generalized P\"oschl-Teller potential}

We next consider the complexification of the generalized P\"oschl-Teller
potential~\cite{cooper}, namely
\begin{equation}
  V(x) = V_1 \cosech^2\tau - V_2 \cosech\tau \coth\tau, \qquad \tau = x - c - {\rm i}
  \gamma, \qquad V_1 > - \case{1}{4}, \qquad V_2 \ne 0.  \label{eq:PT}
\end{equation}
It is easy to recognize (\ref{eq:PT}) to belong to class~II defined in Eq.~(\ref{eq:VmII}).
Note that the above potential is PT-symmetric as well as P-pseudo-Hermitian. Comparing
with (\ref{eq:VmII}), we get
\begin{eqnarray}
  b_R^2 - b_I^2 + m_R^2 - m_I^2 - \case{1}{4} & = & V_1, \label{eq:PT-eq1} \\
  b_R b_I + m_R m_I & = & 0, \\
  2(m_R b_R - m_I b_I) & = & V_2, \\
  m_R b_I + m_I b_R  & = & 0. \label{eq:PT-eq4}  
\end{eqnarray}
This time there is no reason to assume that $b_I \ne 0$, since the presence of $\gamma
\ne 0$ in the generators (\ref{eq:J}) ensures that we are dealing with sl(2, \C).\par
%
%-----------------------------------------------------------------------------------------------------------
%
On successively considering the cases where $b_I = 0$ or $b_I \ne 0$ and proceeding as
in the previous section, we are led to the two following sets of solutions of
Eqs.~(\ref{eq:PT-eq1}) -- (\ref{eq:PT-eq4}):
\begin{eqnarray}
  b_R & = & \case{1}{2} \nu \left(\sqrt{V_1 + \case{1}{4} + |V_2|} - \epsilon
        \sqrt{V_1 + \case{1}{4} - |V_2|}\right), \qquad b_I = 0, \nonumber \\
  m_R & = & \case{1}{2} \left(\sqrt{V_1 + \case{1}{4} + |V_2|} + \epsilon
        \sqrt{V_1 + \case{1}{4} - |V_2|}\right), \qquad m_I = 0, \qquad \epsilon = \pm 1,
        \label{eq:PT-sol1}
\end{eqnarray}
provided $|V_2| \le V_1 + \frac{1}{4}$ and $\sqrt{V_1 + \frac{1}{4} + |V_2|} + \epsilon
\sqrt{V_1 + \frac{1}{4} - |V_2|} > 1$, and
\begin{eqnarray}
  b_R & = & \case{1}{2} \nu \sqrt{|V_2| + V_1 + \case{1}{4}}, \qquad b_I = -
       \case{1}{2} \nu \epsilon \sqrt{|V_2| - V_1 - \case{1}{4}}, \nonumber \\
  m_R & = & \case{1}{2} \sqrt{|V_2| + V_1 + \case{1}{4}}, \qquad m_I =  \case{1}{2}
       \epsilon \sqrt{|V_2| - V_1 - \case{1}{4}}, \qquad \epsilon = \pm 1,
       \label{eq:PT-sol2} 
\end{eqnarray}
provided $|V_2| > V_1 + \frac{1}{4}$ and $|V_2| + V_1 + \case{1}{4} > 1$. In both
cases, $\nu$ denotes the sign of $V_2$.\par
%
%---------------------------------------------------------------------------------------------------------
%
Comparison with Eq.~(\ref{eq:E}) shows that the first type solutions (\ref{eq:PT-sol1})
lead to two series of real eigenvalues
\begin{equation}
  E_{n,\epsilon} = - \left[\case{1}{2} \left(\sqrt{V_1 + \case{1}{4} + |V_2|} + \epsilon
  \sqrt{V_1 + \case{1}{4} - |V_2|}\right) - n - \case{1}{2}\right]^2, \qquad \epsilon =
  \pm 1,  \label{eq:PT-E1}
\end{equation}
while the second type solutions (\ref{eq:PT-sol2}) correspond to a series of
complex-conjugate pairs of eigenvalues
\begin{equation}
  E_{n,\epsilon} = - \left[\case{1}{2}\left(\sqrt{|V_2| + V_1 + \case{1}{4}} +
  {\rm i} \epsilon \sqrt{|V_2| - V_1 - \case{1}{4}}\right) - n - \case{1}{2}\right]^2,
  \qquad \epsilon = \pm 1.
\end{equation}
In the former (resp.\ latter) case, it can be shown that $n$ varies in the range $n=0$, 1,
2,~\ldots ${}< \case{1}{2} \left(\sqrt{V_1 + \case{1}{4} + |V_2|} + \epsilon
\sqrt{V_1 + \case{1}{4} - |V_2|} - 1\right)$ [resp.\ $n=0$, 1, 2,~\ldots ${}<
\case{1}{2}\left(\sqrt{|V_2| + V_1 + \case{1}{4}} - 1\right)$].\par
%
%---------------------------------------------------------------------------------------------------
%
{}For increasing values of $|V_2|$, we observe a phenomenon entirely similar to that
already noted for the complexified Scarf~II potential: disappearance of the real
eigenvalues and simultaneous appearance of complex-conjugate ones at the threshold
$|V_2| = V_1 + \frac{1}{4}$. In this case, however, only partial results were reported in
the literature. In Ref.~\cite{bagchi00a}, we obtained the two series of real eigenvalues
(\ref{eq:PT-E1}) using the parametrization $V_1 = B^2 + A(A+1)$, $V_2 = B(2A+1)$,
with $A$ and $B$ real, so that the condition $|V_2| \le V_1 + \frac{1}{4}$ is
automatically satisfied. Furthermore, L\'evai and Znojil considered both the
real~\cite{levai00} and the complex~\cite{levai01b} eigenvalues in a parametrization
$V_1 = \frac{1}{4}[2(\alpha^2 + \beta^2) - 1]$, $V_2 = \frac{1}{2}(\beta^2 -
\alpha^2)$, wherein $\alpha$ and $\beta$ are real or one of them is real and the other
imaginary, respectively. Their results, however, disagree with ours in both cases.\par 
%
%===========================================================
%
\section{Complexified Morse potential}

The potential
\begin{equation}
  V(x) = (V_{1R} + {\rm i}V_{1I}) e^{-2x} - (V_{2R} + {\rm i}V_{2I}) e^{-x}, \qquad
  V_{1R}, V_{1I}, V_{2R}, V_{2I} \in \R,  \label{eq:Morse} 
\end{equation}
is the most general potential of class~III for the upper sign choice in Eq.~(\ref{eq:VmIII})
and is a complexification of the standard Morse potential~\cite{cooper}. Comparison with
Eq.~(\ref{eq:VmIII}) shows that
\begin{eqnarray}
  b_R^2 - b_I^2 & = & V_{1R},  \label{eq:Morse-eq1} \\
  2b_R b_I & = & V_{1I},  \label{eq:Morse-eq2} \\
  2(m_R b_R - m_I b_I) & = & V_{2R},  \label{eq:Morse-eq3} \\
  2(m_R b_I + m_I b_R) & = & V_{2I},  \label{eq:Morse-eq4}
\end{eqnarray}
where we may assume $b_I \ne 0$.\par 
%
%-------------------------------------------------------------------------------------------------------
%
On solving Eq.~(\ref{eq:Morse-eq2}) for $b_R$ and inserting the result into
Eq.~(\ref{eq:Morse-eq1}), we get a quadratic equation for $b_I^2$, of which we only
keep the real positive solutions. The results for $b_R$ and $b_I$ read
\begin{equation}
  b_R = \case{1}{\sqrt{2}} \epsilon_I \nu (V_{1R} + \Delta)^{1/2}, \quad b_I =
  \case{1}{\sqrt{2}} \epsilon_I (- V_{1R} + \Delta)^{1/2}, \quad \Delta = 
  \sqrt{V_{1R}^2 + V_{1I}^2}, \quad \epsilon_I = \pm 1, \label{eq:Morse-inter} 
\end{equation}
where $V_{1I} \ne 0$ if $V_{1R} \ge 0$ and $\nu$ denotes the sign of $V_{1I}$. On
introducing Eq.~(\ref{eq:Morse-inter}) into Eqs.~(\ref{eq:Morse-eq3}) and
(\ref{eq:Morse-eq4}) and solving for $m_R$ and $m_I$, we then obtain
\begin{eqnarray}
  m_R & = & \frac{\epsilon_I \nu}{2\sqrt{2} \Delta} \left[(V_{1R} + \Delta)^{1/2}
        V_{2R} + \nu (- V_{1R} + \Delta)^{1/2} V_{2I}\right], \\
  m_I & = & \frac{\epsilon_I \nu}{2\sqrt{2} \Delta} \left[(V_{1R} + \Delta)^{1/2}
        V_{2I} - \nu (- V_{1R} + \Delta)^{1/2} V_{2R}\right].
\end{eqnarray}
\par
%
%-------------------------------------------------------------------------------------------------------
%
{}From the regularity conditions $b_R > 0$ and $m_R > \frac{1}{2}$ of
$\psi^{(m)}_0(x)$, given in Eq.~(\ref{eq:psiIII}), it follows that we must choose
$\epsilon_I = \nu$, $V_{1I} \ne 0$ if $V_{1R} < 0$, and
\begin{equation}
  (V_{1R} + \Delta)^{1/2} V_{2R} + \nu (- V_{1R} + \Delta)^{1/2} V_{2I} > \sqrt{2}
  \Delta.  \label{eq:Morse-cond1}
\end{equation}
We conclude that $V_{1I} \ne 0$ must hold for any value of $V_{1R}$.\par
%
%-------------------------------------------------------------------------------------------------------
%
Real eigenvalues are associated with $m_I = 0$ and therefore occur whenever the
condition
\begin{equation}
  (V_{1R} + \Delta)^{1/2} V_{2I} = \nu (- V_{1R} + \Delta)^{1/2} V_{2R}
  \label{eq:Morse-cond2}
\end{equation}
is satisfied. In such a case, $V_{2I}$ can be expressed in terms of $V_{1R}$, $V_{1I}$,
and $V_{2R}$, so that the real eigenvalues are given by
\begin{equation}
  E_n = - \left[\frac{V_{2R}}{\sqrt{2}|V_{1I}|} (- V_{1R} + \Delta)^{1/2} - n -
  \frac{1}{2}\right]^2.
\end{equation}
It can be shown that regular eigenfunctions correspond to $n=0$, 1, 2,~\ldots ${}<
(V_{2R}/\sqrt{2}|V_{1I}|) (- V_{1R} + \Delta)^{1/2} - \frac{1}{2}$.\par
%
%----------------------------------------------------------------------------------------------------
%
{}Furthermore, when condition (\ref{eq:Morse-cond2}) is not fulfilled but condition
(\ref{eq:Morse-cond1}) holds, we get complex eigenvalues associated with regular
eigenfunctions,
\begin{equation}
  E_n = - \left\{\frac{1}{2\sqrt{2} \Delta} \left[(V_{1R} + \Delta)^{1/2} - {\rm i} \nu
  (- V_{1R} + \Delta)^{1/2}\right] (V_{2R} + {\rm i}V_{2I}) - n - \frac{1}{2}\right\}^2,
\end{equation}
where $n = 0$, 1, 2,~\ldots ${}< \frac{1}{2\sqrt{2} \Delta} \left[(V_{1R} + 
\Delta)^{1/2} V_{2R} + \nu (- V_{1R} + \Delta)^{1/2} V_{2I}\right] -
\frac{1}{2}$.\par
%
%------------------------------------------------------------------------------------------------------
% 
It should be stressed that contrary to what happens for the two previous examples, here
the real eigenvalues, belonging to a single series, only occur for a special value of the
parameter $V_{2I}$, while the complex eigenvalues, which do not appear in
complex-conjugate pairs (since $E_n^*$ corresponds to $V^*(x)$), are obtained for
generic values of $V_{2I}$.\par
%
%-----------------------------------------------------------------------------------------------------
%
To interprete such results, it is worth choosing the parametrization $V_{1R} = A^2 -
B^2$, $V_{1I} = 2AB$, $V_{2R} = \gamma A$, $V_{2I} = \delta B$, where $A$, $B$,
$\gamma$, $\delta$ are real, $A>0$, and $B \ne 0$. The complexified Morse potential
(\ref{eq:Morse}) can then be expressed as 
\begin{equation}
  V(x) = (A + {\rm i}B)^2 e^{-2x} - (2C+1) (A + {\rm i}B) e^{-x}, \qquad C = 
  \frac{(\gamma-1) A + {\rm i} (\delta-1) B}{2(A + {\rm i} B)}.  \label{eq:Morse-bis}
\end{equation}
Its (real or complex) eigenvalues can be written in a unified way as $E_n = - (C-n)^2$,
while the regularity condition (\ref{eq:Morse-cond1}) amounts to $(\gamma-1) A^2 +
(\delta-1) B^2 > 0$.\par
%
%---------------------------------------------------------------------------------------------------------
%
{}For $\delta = \gamma > 1$, and therefore $C = \frac{1}{2} (\gamma-1) \in \R^+$,
the potential (\ref{eq:Morse-bis}) coincides with that considered in our previous
work~\cite{bagchi00a}. Such a potential was shown to be pseudo-Hermitian under
imaginary shift of the coordinate~\cite{ahmed01a}. We confirm here that it has only real
eigenvalues corresponding to $n=0$, 1, 2,~\ldots ${}< C$, thus exhibiting no symmetry
breaking over the whole parameter range. For the values of $\delta$ different from
$\gamma$, the potential indeed fails to be pseudo-Hermitian. In such a case, $C$ is
complex as well as the eigenvalues. The eigenfunctions associated with $n=0$, 1,
2,~\ldots ${}< {\rm Re}\,C$ are however regular. The existence of regular eigenfunctions
with complex energies for general complex potentials is a phenomenon that has been
known for some time (see e.g.~\cite{baye}).\par
%
%=========================================================
%
\section{Conclusion}

In the present Letter, we have shown that complex Lie algebras (in particular sl(2, \C))
provide us with an elegant tool to easily determine both real and complex eigenvalues of
non-Hermitian Hamiltonians, corresponding to regular eigenfunctions. For such a purpose,
it has been essential to extend the scope of our previous work~\cite{bagchi00a} to those
Lie algebra irreducible representations that remain nonunitary in the real algebra limit
(namely those with $k_I \ne 0$).\par
%
%---------------------------------------------------------------------------------------------------------
%
We have illustrated our method by deriving the real and complex eigenvalues of the
PT-symmetric complexified Scarf~II potential, previously determined by other
means~\cite{ahmed01b}. In addition, we have established similar results for the
PT-symmetric generalized P\"oschl-Teller potential, for which only partial results were
available~\cite{bagchi00a, levai00, levai01b}. We have shown that in both cases
symmetry breaking occurs for a given value of one of the potential parameters.\par
%
%--------------------------------------------------------------------------------------------------------
%
{}Finally, we have considered a more general form of the complexified Morse potential
than that previously studied~\cite{bagchi00a, solombrino, ahmed01a}. For a special value
of one of its parameters, our potential reduces to the former one and becomes
pseudo-Hermitian under imaginary shift of the coordinate. We have proved that here no
symmetry breaking occurs, the complex eigenvalues being associated with
non-pseudo-Hermitian Hamiltonians.\par
%
%===========================================================
%

\end{document}